\documentclass[twocolumn,prb,showpacs]{revtex4}
\usepackage{graphicx}
\usepackage{dcolumn}
\usepackage{bm}

\begin{document}

\title{Giant Dielectric Permittivity of Electron-Doped Manganite Thin Films,
Ca$_{1-x}$La$_x$MnO$_3$ ($0\leq x\leq 0.03$)}
\author{J. L. Cohn}
\affiliation{Department of Physics, University of Miami, Coral
Gables, Florida 33124}
\author{M. Peterca$^{\dagger}$}
\affiliation{Department of Physics, University of Miami, Coral
Gables, Florida 33124}
\author{J. J. Neumeier}
\affiliation{Department of Physics, Montana State University,
Bozeman, Montana 59717}

\begin{abstract}
A giant low-frequency, in-plane dielectric constant, $\epsilon\sim
10^6$, for epitaxial thin films of Ca$_{1-x}$La$_x$MnO$_3$ ($x\leq
0.03$) was observed over a broad temperature range, 4K~$\leq
T\leq$~300K.  This phenomenon is attributed to an internal
barrier-layer capacitor (IBLC) structure, with Schottky contacts
between semiconducting grains. The room-temperature
$\epsilon$ increases substantially with electron (La) doping, consistent
with a simple model for IBLCs. The
measured values of $\epsilon$ exceed those of conventional
two-phase IBLC materials based on (Ba,Sr)TiO$_3$ as well as
recently discovered CaCu$_3$Ti$_4$O$_{12}$  and (Li,Ti) doped NiO.
\end{abstract}

\maketitle

\section{\label{sec:Intro} INTRODUCTION}

High-permittivity dielectric materials play an important role in
electroceramic devices such as capacitors and memories.  Recent
reports of giant permittivity have directed considerable attention
to several new material systems: non-ferroelectric
CaCu$_3$Ti$_4$O$_{12}$,\cite{CCTO} percolative BaTiO$_3$-Ni
composites,\cite{Percol} and (Li,Ti)-doped NiO.\cite{NiO} Of
particular interest for applications is the weakly temperature
dependent permittivity of these materials near room temperature.
The enormous dielectric constant of these materials, $\epsilon\sim
10^5$, appears to be\cite{CCTOisIBLC} a consequence of an internal
barrier-layer capacitor (IBLC) structure, composed of insulating
layers between semiconducting grains.  IBLCs with effective
$\epsilon\sim 10^5$ based on (BaSr)TiO$_3$ are well
known,\cite{IBLCs} but their usefulness is limited by a strong
frequency and temperature dependence of their dielectric constant.
Thus the newly discovered materials suggest that increases in
$\epsilon$ values and/or simplification of processing could yield
new and useful IBLC materials.

Here we report on giant values of the effective dielectric constant,
$\epsilon\sim 10^6$, observed at low frequencies ($f\leq 100$~kHz)
for thin films of the electron-doped manganite,
Ca$_{1-x}$La$_x$MnO$_3$ ($0\leq x\leq 0.03$).  The large and weakly
$T$- and $f$-dependent $\epsilon$ near room temperature is
attributed to an IBLC microstructure, comprising a network of depletion
layers between semiconducting grains. $\epsilon$ is enhanced by
electron doping via La substitution for Ca or oxygen reduction.

\section{\label{sec:Expt} EXPERIMENTAL METHODS}

Polycrystalline targets of Ca$_{1-x}$La$_x$MnO$_3$ ($x\leq 0.03$)
were prepared by solid-state reaction as described
previously.\cite{NeumeierCohn} Powder x-ray diffraction revealed
no secondary phases and iodometric titration, to measure the
average Mn valance, indicated an oxygen content for all targets
within the range 3.00$\pm$0.01.  Thin films were grown by pulsed
laser deposition using a 248 nm KrF excimer laser, with energy
density $\sim 0.8-1.2$ J/cm$^2$, pulse repetition rate 10 Hz, and
target-substrate distance, 4 cm. The films were deposited on
LaAlO$_3$ (LAO) substrates of [100] orientation, with substrate
temperature 750 $^{\circ}$C and oxygen pressure 200~mTorr.
Following the depositions, the chamber was filled to 700~Torr
oxygen, held at 500~$^{\circ}$C for 30 min., and subsequently
cooled to room temperature. Film thicknesses were $\sim
50-170~$nm. X-ray diffraction (XRD) indicated epitaxial growth of
the pseudocubic perovskite for all films, with lattice parameter,
$a=3.72{\rm\AA}$ for CaMnO$_3$.  The full widths at half maximum
of the (200) film reflections were typically 0.4$^{\circ}$.
Scanning electron microscopy indicated an average grain size of
0.5-1$\mu$m.
\begin{figure}
\vglue -.1in
\includegraphics[width = 5.in]{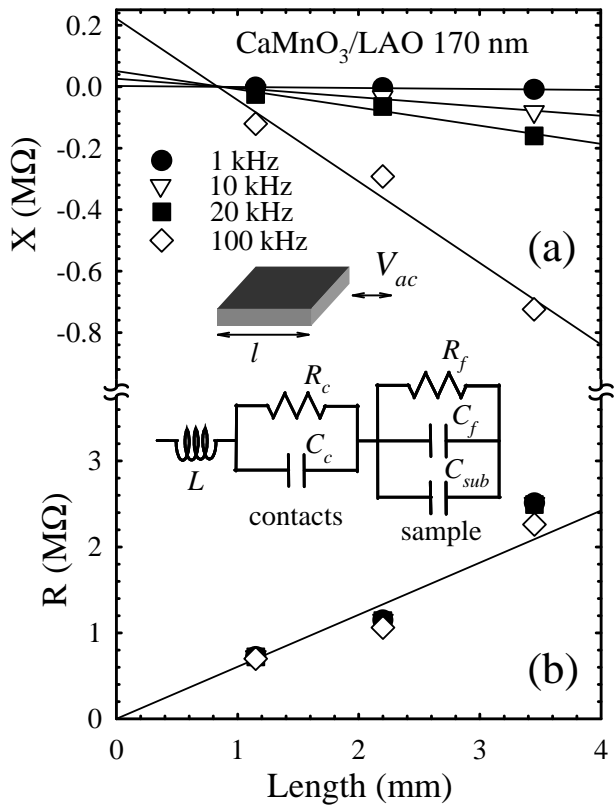}%
\vskip -.4in
\caption{(a) $X(l)$ and (b) $R(l)$ for a 170-nm
CaMnO$_3$/LAO film at room temperature. The inset shows the
contact configuration and equivalent
circuit.\label{LengthDependence}}
\end{figure}
Impedance measurements were performed with a Hewlett-Packard model
4263B LCR meter in the frequency range 100Hz to 100kHz. A model
HP16034E test fixture was used at room temperature, and a
coaxial-lead, four-terminal pair arrangement in a cryostat for low
temperatures. Silver paint electrodes, annealed at 300$^{\circ}$C,
were applied on opposing edges of the specimens so that the ac
voltage (0.2V for all measurements) was applied in the film plane.
In this configuration, the film and substrate capacitances are
additive; the equivalent circuit for contacts, film and substrate
is shown in the inset of Fig.~\ref{LengthDependence}.  A contact
capacitance can lead to apparently large
values\cite{LunkenheimerEprint} of $\epsilon$ and thus great care
is required to distinguish the true response of the sample.  To
address this issue, contact contributions to the impedance were
eliminated at room temperature for some films by measuring the
length dependence of the impedance (film+substrate), $Z=R+jX$, on
a series of films deposited simultaneously onto pre-cut substrates
of different length, $l$. Since the contact capacitance and
resistance should be independent of $l$, the measured reactance
and resistance should be linear in $l$, $X=X_0-\beta l$ and
$R=R_0+\beta^{\prime} l$, respectively. $X_0=\omega L-\omega R_C^2
C_C/[1+(\omega R_C C_C)^2]$ is a constant determined by a small
serial inductance, $L$ (also assumed independent of
$l$),\cite{NoteonInductance} and the capacitive reactance of the
contacts. These relations are followed well by the raw data
(Fig.~\ref{LengthDependence}). The dielectric constant of the film
was computed as,
\begin{eqnarray}
\epsilon_f={1\over \omega A_f\epsilon_0}{\beta\over \beta^2+\beta^{\prime~2}}-\epsilon_{sub}{A_{sub}\over A_f,}
\label{epsFilm}
\end{eqnarray}
\noindent where $\omega=2\pi f$ is the angular frequency of the
applied voltage, $A_f$ ($A_{sub}$) is the film (substrate) contact
area and $\epsilon_{sub}$ is the substrate dielectric constant.
The latter, determined from a similar analysis of $Z(l)$ for blank
substrates, was $\epsilon_{sub}=24$ in good agreement with
published values.\cite{SubstrateEps} Uncertainty in $\epsilon_f$
determined from Eq.~(1), arising principally from scatter in the
data, was typically $\pm 20$\%.  Values for $\epsilon_f$
determined from the length dependence analysis were consistently
larger than those determined from direct measurements, indicating
a predominance of the inductive reactance over the capacitive
reactance of the contacts (i.e. $X_0>0$). The $\epsilon(T)$ data
presented below from direct measurements thus underestimate the
true magnitude. The uncertainty in $\epsilon$ for the direct
measurements near room temperature (where $C_f\gg C_{sub}$) is
given by that of the film thickness ($\pm 10\%$). At the lowest
temperatures, where $C_f\ll C_{sub}$, the uncertainty in
$\epsilon$ is $\pm50\%$, largely due to a $\pm5\%$ uncertainty in
$C_{sub}$.  For all measurements, the impedance was independent of
ac voltage (50 mV-1 V) and applied DC bias (0-2 V).

\section{\label{sec:Results} RESULTS AND DISCUSSION}

\begin{figure}
\vglue -.1in
\includegraphics[width = 4.in]{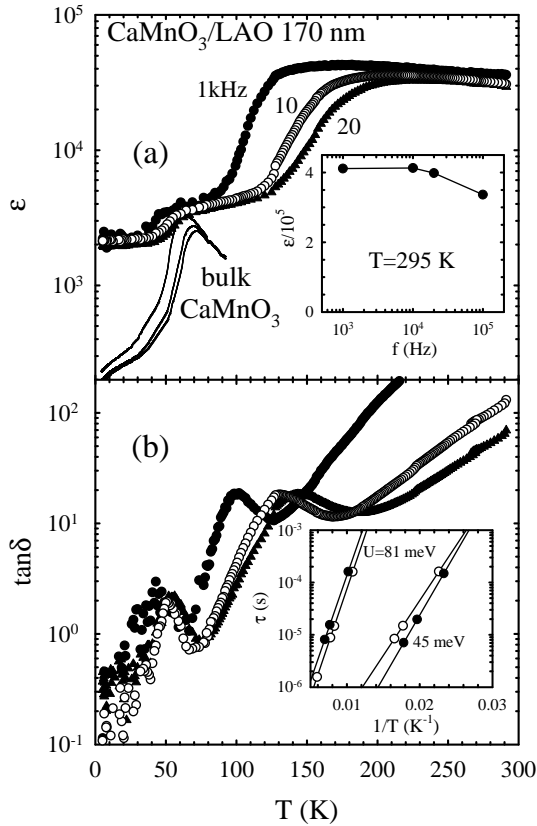}%
\vglue .5in
\caption{a) $\epsilon (T)$ and (b) $\tan\delta (T)$ at three
frequencies for a 170-nm CaMnO$_3$/LAO film.  The solid curves in
(a) are for bulk CaMnO$_3$ at the same frequencies.  The inset in
(a) shows $\epsilon(f)$ at room temperature for a second piece of
the same film.  The inset in (b) shows Arrhenius behavior of
relaxation times determined from peaks in $\tan\delta$ for both
pieces of the film and corresponding average activation energies.
\label{CMOEpsvsT}}
\end{figure}

The impedance of the target materials was measured in separate
experiments.  CaMnO$_3$ is an antiferomagnetic ($T_N\approx 125$K)
semiconductor with a small electron density associated with native
defects, particularly oxygen vacancies, that yield a substantial
room temperature conductivity, $\sigma\sim 100\
\Omega^{-1}$m$^{-1}$. Electron doping via La substitution for Ca
further enhances $\sigma$, especially at low
$T$.\cite{NeumeierCohn}  $\epsilon$ could be determined reliably
only at $T\lesssim 100$ K where the capacitive reactance was
sufficiently large ($\gtrsim 0.1\Omega$). $\epsilon$ was
describable as a sum of a constant term $\sim 25-50$ (the
high-frequency and lattice terms) and a dipolar contribution
associated with hopping charge carriers.\cite{Jonscher} The latter
gives rise to steps in $\epsilon(T)$ which occur at lower $T$ with
decreasing frequency, as shown for CaMnO$_3$ in
Fig.~\ref{CMOEpsvsT}~(a) (solid curves). This indicates a
relaxation process associated with thermal activation of localized
charge carriers, characterized by a relaxation time,
$\tau=\tau_0\exp(U/k_BT)$. Analyzing the maxima in $d\epsilon/dT$
(corresponding to $\omega\tau=1$) we find $U=103$~meV and
$\tau_0=7.5\times10^{-14}$~s. These parameters are typical of
polaronic relaxation in LaMnO$_3$\cite{Lunkenheimer} and other
perovskites.\cite{Bidault}  For comparison, $U=54$ meV,
$\tau_0=8.4\times 10^{-10}$~s were reported for single-crystal
CaCu$_3$Ti$_4$O$_{12}$,\cite{Homes} and $U=313$ meV,
$\tau_0=8.5\times10^{-13}$~s for (Li,Ti)-doped NiO.\cite{NiO}
Relevant to the subsequent discussion of films, it was found that
increasing the electron density, either through oxygen reduction
or La doping, resulted in a decrease (increase) in $U\ (\tau_0)$.
Further details of measurements on the polycrystalline bulk
materials will be presented elsewhere.\cite{CohnUnpub}

\begin{figure}
\vglue .1in
\includegraphics[width = 3.9in]{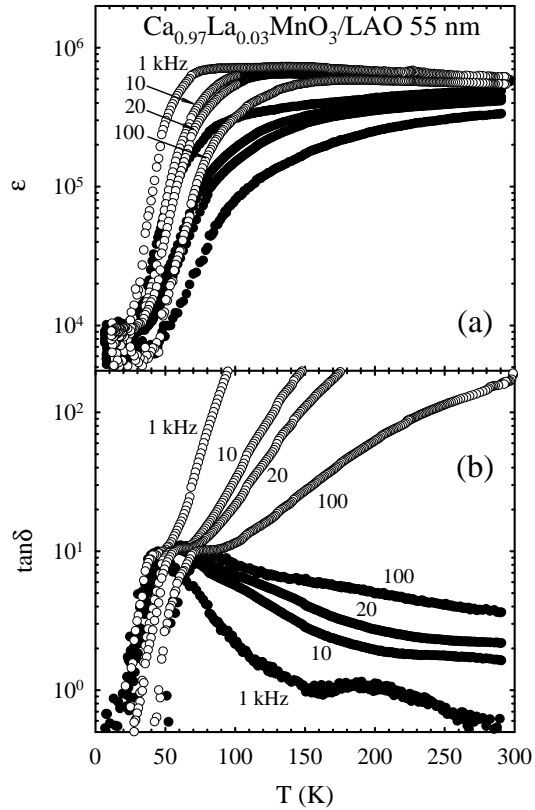}%
\vskip .4in
\caption{(a) $\epsilon (T)$ and (b) $\tan\delta (T)$ at several
frequencies for a 55-nm Ca$_{0.97}$La$_{0.03}$MnO$_3$/LAO film,
as-prepared (open symbols) and ``aged'' for 3 months in air at
room temperature (closed symbols).\label{EpsvsT}}
\end{figure}

Figure~\ref{CMOEpsvsT}~(a) shows $\epsilon (T)$ at three
frequencies ($\leq 20$ kHz) for a 170-nm film of CaMnO$_3$. The
inset shows $\epsilon(f)$ at room temperature for a second piece
of the same film measured to 100~kHz.  At low temperatures,
$\epsilon$ has a magnitude comparable to the bulk material and
small steps occur at temperatures similar to those at which the
bulk material exhibits dipolar relaxation. With increasing
temperature a second dispersive step is observed near 200 K, with
$\epsilon$ increasing by an order of magnitude. For $T>200$ K,
$\epsilon\sim 3-4\times 10^4$ with weak temperature and frequency
dependencies.  Both steps in $\epsilon$ appear as maxima in the
dielectric loss [$\tan\delta$, Fig.~\ref{CMOEpsvsT}~(b)]. The
appearance of the high-temperature step is typical of IBLC
materials where insulating barriers separate semiconducting grain
interiors. Such a system can be modelled as two $RC$ circuits in
series, one for the grain interiors and one for the grain-boundary
response.\cite{CCTOisIBLC} The grain-boundary $RC$ time constant,
$\tau=R_{gb}C_{gb}$, gives rise to a second relaxation process
that is thermally activated via the behavior of $R_{gb}$. From the
temperatures of the maxima in $\tan\delta$ measured for two pieces
of this film we compute [inset, Fig.~\ref{CMOEpsvsT} (b)] average
values for activation energies $U=45\pm 6\ [81\pm 2]$~meV and
$\tau_0=(1.6\pm 1.2)\times10^{-9}$~s [($9.4\pm
3.5)\times10^{-9}$~s] for the grain interior [grain-boundary]
relaxations. The values of $U$ and $\tau_0$ inferred for the grain
interiors are significantly smaller and larger, respectively, than
the corresponding values determined for the targets. This
indicates that the grains of the film are more oxygen deficient,
and suggests oxygen is lost during deposition.  Consistent with
this conclusion, the films have\cite{CohnUnpub} $T_N\approx 200$K
in accord with magnetic measurements on oxygen deficient, bulk
CaMnO$_{3-\delta}$.\cite{Briatico}
Figure~\ref{EpsvsT} shows $\epsilon (T)$ and $\tan\delta (T)$
for a 55-nm film grown from a Ca$_{.97}$La$_{0.03}$MnO$_3$ target.
The qualitative features of the data for the as-prepared specimen
(open symbols) are similar to that of the CaMnO$_3$ film except
that $\epsilon$ is substantially larger, $\sim 6\times 10^5$ at
300 K and the lower temperature step in $\epsilon$ is absent. The
higher carrier density of the La-doped material suppresses the
carrier freeze-out responsible for the low-$T$ relaxation to
temperatures below our measurement range.\cite{CohnUnpub} A
greater dc conductivity is responsible for the sharp increase of
$\tan\delta$ at higher temperatures. The grain-boundary relaxation
is evident as weak maxima or plateaus in $\tan\delta$. Analysis of
the relaxation yields, $U=32$~meV and $\tau_0=1.1\times
10^{-8}$~s.
After sitting in ambient conditions for three months, $\epsilon$
at 300 K for the same specimen (solid symbols) decreased by $\sim
20-30$\%, but $\tan\delta$ had decreased dramatically; by more
than four orders of magnitude at 1 kHz and nearly two orders of
magnitude at 100 kHz (Fig.~\ref{EpsvsT}).  At lower temperatures,
$\tan\delta$ exhibits well-defined maxima, and below these maxima
matches the data in the as-prepared state.  Thus much of the loss
near room temperature in the as-prepared films is associated with
dc conduction that is substantially suppressed with ``aging''.

As for other IBLCs composed of compound semiconductors, our data
can be understood by considering the films to be random arrays of
close-packed, electrically-active semiconducting grains.
Boundaries between grains contain an interface charge $Q_i$, and
are adequately described as double Schottky
barriers\cite{GreuterBlatter} with capacitance,
$C_{DSB}=\epsilon\epsilon_0A/2x_0$, where $A$ is the area of
contact between grains, $\epsilon$ is the bulk semiconductor
dielectric constant, $x_0=Q_i/2N_0$ is the depletion layer width
on either side of the grain boundary (assumed symmetric), and
$N_0$ is the donor density in the bulk semiconducting grains.
Since the grain diameter $D$ is typically $\gg x_0$, the
``brickwork'' model\cite{IBLCs} can be applied to the array of
grains, such that the effective dielectric constant is given
approximately as,
$\epsilon_{eff}\approx (D/2x_0)\epsilon=\epsilon D N_0/Q_i$.

\begin{figure}
\includegraphics[width = 3.7in]{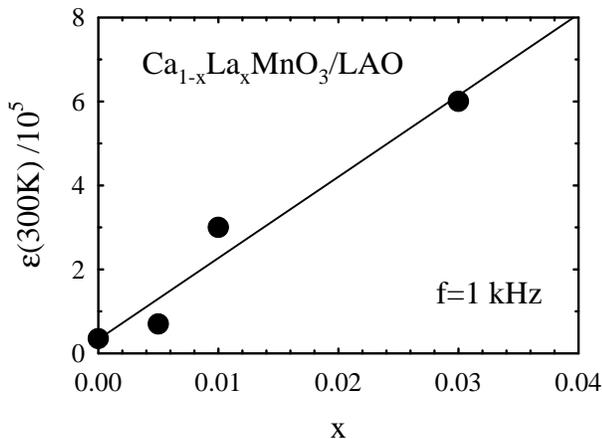}%
\vglue -2in
\caption{Doping dependence of the room temperature $\epsilon$ at 1
kHz.\label{EpsvsX}}
\end{figure}

With increased electron (La) doping,
$\epsilon$ for the films increases considerably (Fig.~\ref{EpsvsX}).
This data allows for a test of the simple model above.  Reasonably assuming that
the oxygen vacancy concentration ($y$) for the films is independent of $x$, we
compute the slope, $d\epsilon_{eff}/dx$, by making the
substitution $N_0=(x+2y)/V_{fu}$,
\begin{eqnarray}
{d\epsilon_{eff}\over dx}={\epsilon D \over Q_iV_{fu}}
\label{DepsDx}
\end{eqnarray}
where $V_{fu}=51{\rm\AA^3}$ is the volume per formula unit.\cite{Ling}
The line in Fig.~\ref{EpsvsX} yields $d\epsilon_{eff}/dx\approx 2\times 10^7$.
Using $\epsilon\sim 10^3$ (Fig.~\ref{CMOEpsvsT} and Ref.~\onlinecite{CohnUnpub})
and $D=0.5\ \mu{\rm m}$ (from scanning electron microscopy),
we compute an interface charge density of
$Q_i\simeq 5\times 10^{17} {\rm m}^{-2}$.  This value is typical of those found
for a variety of compound semiconductors.\cite{GreuterBlatter}  That $\epsilon_{eff}$ is
approximately linear in $x$ implies that variations in $\epsilon$ and $Q_i$ with doping
are either negligible or cancel.

The aging effect in the films is most likely associated with oxidation at
grain surfaces. This hypothesis is supported by preliminary oxygen
annealing studies on a 55 nm CMO film. $\tan\delta$ was reduced by
a factor of two with no measurable change in $\epsilon$ following
a 15-hour, flowing-oxygen anneal at 700$^{\circ}$C.  A small
decrease in $\epsilon$ for aged films is consistent with a small
increase in $Q_i$.

Finally, we note that the large values of $\epsilon$ reported here are
not limited to manganite films with compositions near 100\%
Ca.  We have measured $\epsilon\sim 10^4$ at room temperature for a
55 nm film grown under identical conditions from a La$_{0.7}$Ca$_{0.3}$MnO$_3$
target (a colossal magnetoresistance composition).  Films of other insulating
compositions with the appropriate microstructure may also possess a large
effective $\epsilon$.
\break

\section{CONCLUSIONS}

In summary, thin films of Ca$_{1-x}$La$_x$MnO$_3$ (0$\leq
x\leq$0.03) have been found to exhibit giant low-frequency
dielectric constants, $\epsilon\sim 10^6$, that are weakly
temperature and frequency dependent near room temperature.  These
enormous values are attributed to a barrier-layer capacitor
microstructure produced during deposition and subsequent exposure
to air by oxidation of grain boundary regions which form an
insulating shell on semiconducting grains.  $\epsilon$ increases
with charge-carrier doping, consistent with a reduction in the
depletion-layer width at grain boundary contacts and a nearly
doping-independent surface charge. Though lower dielectric losses
will be required for
applications ($\tan\delta\leq 0.05$ is desirable), the
considerable reduction of $\tan\delta$ upon aging or oxygen
annealing with little decrease in $\epsilon$ motivates further
investigations of processing conditions.

\section*{ACKNOWLEDGMENTS}

The authors gratefully acknowledge experimental assistance from
Dr. B. Zawilski.  The work at the Univ. of Miami was supported, in
part, by NSF Grant No.'s DMR-9504213 and DMR-0072276, and at
Montana State Univ. by Grant No. DMR-0301166.

\noindent
$^{\dagger}$ present address: Physics Department, University of Pennsylvania,
Philadelphia, PA

\end{document}